\documentclass[pdflatex,sn-mathphys-num]{sn-jnl}

\usepackage{graphicx}%
\usepackage{multirow}%
\usepackage{amsmath,amssymb,amsfonts}%
\usepackage{amsthm}%
\usepackage{mathrsfs}%
\usepackage[title]{appendix}%
\usepackage{xcolor}%
\usepackage{textcomp}%
\usepackage{manyfoot}%
\usepackage{booktabs}%
\usepackage{newunicodechar}
\newunicodechar{⋆}{\ensuremath{\star}}
\usepackage{subcaption}%
\usepackage{algorithm}%
\usepackage{algorithmicx}%
\usepackage{algpseudocode}%
\usepackage{listings}%

\theoremstyle{thmstyleone}%

\theoremstyle{thmstyletwo}%

\theoremstyle{thmstylethree}%

\begin{document}

\title[Article Title]{Equilibrium \(\mathrm{H}_2^+/\mathrm{H}^-\) Abundance Ratios in MARCS Atmosphere Models}


\author*[1]{\fnm{Aiganym} \sur{Sarsembayeva}}\email{sarsembaeva.a@kaznu.kz}

\author[2,3]{\fnm{Hernando} \sur{Quevedo}}\email{quevedo@nucleares.unam.mx}
\equalcont{These authors contributed equally to this work.}

\affil*[1]{\orgdiv{Department of Theoretical and Nuclear Physics}, \orgname{Farabi University}, \orgaddress{\street{71 Al-Farabi Ave.}, \city{Almaty}, \postcode{050040}, \country{Kazakhstan}}}

\affil[2]{\orgdiv{Instituto de Ciencias Nucleares}, \orgname{Universidad Nacional Autónoma de México}, \city{Mexico City}, \postcode{AP 70543}, \country{Mexico}}

\affil[3]{\orgdiv{Dipartimento di Fisica and ICRA}, \orgname{Università di Roma “La Sapienza”}, \orgaddress{\street{Piazzale Aldo Moro 5}, \city{Roma}, \postcode{I-00185}, \country{Italy}}}


\abstract{We investigate the equilibrium abundance ratio of the molecular hydrogen ion 
\(\mathrm{H}_2^+\) and the negative hydrogen ion \(\mathrm{H}^-\) in stellar 
atmospheres using compact Saha-type equilibrium expressions and layer-by-layer 
partial pressures from MARCS model atmospheres. The analysis distinguishes the 
temperature-dependent coefficient ratio 
\(K_{\mathrm{H}_2^+}(T)/K_{\mathrm{H}^-}(T)\) from the full atmospheric abundance 
ratio, which additionally depends on the local ionisation factor 
\(N_{\mathrm{H}^+}/N_e\). The internal partition function 
\(Z_{\mathrm{H}_2^+}(T)\) is included in the \(\mathrm{H}_2^+\) equilibrium 
coefficient. We processed \(51{,}994\) MARCS stellar-atmosphere models, 
corresponding to \(2{,}911{,}664\) atmospheric layers, and computed 
\(N(\mathrm{H}_2^+)/N(\mathrm{H}^-)\) directly from the tabulated equilibrium 
partial pressures, thereby constructing the first systematic map of the equilibrium abundance ratio  
 across the MARCS stellar-atmosphere grid. 
Across the full grid, \(\mathrm{H}_2^+\) exceeds 
\(\mathrm{H}^-\) in \(827{,}350\) layers, corresponding to \(F=0.284\), or 
approximately \(28.4\%\) of all analysed layers. Although \(\mathrm{H}^-\) 
remains dominant in the global logarithmic mean, with 
\(\langle \log_{10}[N(\mathrm{H}_2^+)/N(\mathrm{H}^-)] \rangle=-4.229\), 
\(\mathrm{H}_2^+\) becomes comparable to or more abundant than \(\mathrm{H}^-\) 
in a substantial subset of atmospheric layers. The relative significance of 
\(\mathrm{H}_2^+\) increases toward higher \(T_{\mathrm{eff}}\), lower 
\([\mathrm{Fe}/\mathrm{H}]\), and deeper atmospheric layers. These results show 
that the \(\mathrm{H}_2^+/\mathrm{H}^-\) balance is controlled not by temperature 
alone, but by the combined local thermodynamic and ionisation structure of the 
atmosphere. The reported ratios are equilibrium abundance diagnostics and do not 
imply opacity dominance; a direct opacity assessment requires wavelength-dependent 
cross-sections and radiative-transfer calculations.}




\maketitle

\section{Introduction}\label{sec1}

Chemical and ionisation equilibrium calculations are a fundamental part of stellar-atmosphere modelling \cite{Short2021}. 
At each atmospheric depth, the relative populations of atoms, ions, and molecules determine the local equation of state, the electron pressure, and the continuous and line opacities that shape the emergent spectrum \cite{LynasGray2018}. 
This is particularly important in cool and late-type stellar atmospheres, where molecular formation, ionisation balance, and the electron contribution from metals are strongly coupled \cite{Tsuji1973}. 
Because hydrogen is the most abundant element in stellar photospheres, even hydrogen-bearing species with relatively small number densities can be relevant to the thermodynamic and radiative state of the atmosphere when their equilibrium abundances or radiative cross-sections become favourable under local conditions \cite{atoms11030061}.

Among hydrogen-bearing species, the negative hydrogen ion \(\mathrm{H}^-\) has a special status in stellar-atmosphere physics. 
It is widely recognised as the dominant source of continuous opacity in the visual and near-infrared spectral regions of F-, G-, and K-type stellar atmospheres, including the solar photosphere \cite{atoms11030061, BarklemAmarsi2024, Wishart1979,BellBerrington1987}. 
Its formation is governed by the attachment equilibrium
\begin{equation}
    \mathrm{H} + e^- \rightleftharpoons \mathrm{H}^- ,
    \label{eq:intro_hminus_reaction}
\end{equation}
so that the abundance of \(\mathrm{H}^-\) is directly linked to both the neutral hydrogen density and the free-electron density. 
As a result, the \(\mathrm{H}^-\) population is sensitive to local temperature, gas pressure, electron pressure, metallicity, and atmospheric depth. 
This sensitivity explains why \(\mathrm{H}^-\) has remained a central species in stellar-atmosphere modelling, radiative-transfer calculations, and opacity studies. 
Although \(\mathrm{H}^-\) is commonly treated in local thermodynamic equilibrium (LTE) in one-dimensional model atmospheres, recent work has revisited its statistical equilibrium and found that direct non-LTE effects may reach the percent level in hotter and lower-gravity late-type stars, while being much smaller near solar parameters \cite{BarklemAmarsi2024}. 
This reinforces the need to understand the equilibrium behaviour of \(\mathrm{H}^-\) and its coupling to other hydrogen-bearing species in stratified atmospheres \cite{Short2021}.

The positive molecular hydrogen ion \(\mathrm{H}_2^+\) is another basic component of hydrogen chemistry. 
It is the simplest molecular ion, consisting of two protons bound by a single electron, and is formed through the proton-association equilibrium
\begin{equation}
    \mathrm{H} + \mathrm{H}^+ \rightleftharpoons \mathrm{H}_2^+ .
    \label{eq:intro_h2plus_reaction}
\end{equation}
Owing to its simple structure, \(\mathrm{H}_2^+\) has long served as a benchmark system for quantum-mechanical calculations of molecular structure, photodissociation, and free--free radiative transitions. 
The bound--free photodissociation of \(\mathrm{H}_2^+\) and free--free radiative transitions in the transient \(\mathrm{H}+\mathrm{H}^+\) quasi-molecule contribute to continuous absorption in hydrogen plasmas, and their relative importance has been investigated over wide ranges of wavelength and temperature \cite{Lebedev2000, LebedevPresnyakov2002,Lebedev2003}. 
These studies showed that the contribution of \(\mathrm{H}_2^+\) can become comparable to that of \(\mathrm{H}^-\) under some thermodynamic conditions, although the comparison depends strongly on temperature and wavelength. 
However, such calculations were generally performed for idealised or quasi-equilibrium hydrogen plasmas characterised by a single temperature and simplified ionisation balance, rather than for the depth-dependent pressure and temperature structure of realistic stellar atmospheres.

Reliable equilibrium calculations also require accurate partition functions and equilibrium constants. 
Classical compilations of molecular partition functions and dissociation equilibrium constants have long been used in stellar-atmosphere modelling, and these data have been substantially revised with updated spectroscopic constants and dissociation energies \cite{SauvalTatum1984,Irwin1981, BarklemCollet2016, PopovasJorgensen2016}. 
For molecular species, the internal partition function determines how electronic, vibrational, and rotational states contribute to the equilibrium population. 
This is especially relevant for \(\mathrm{H}_2^+\), because its rovibrational structure enters directly into the equilibrium coefficient that relates the abundance of \(\mathrm{H}_2^+\) to the densities of neutral hydrogen and protons \cite{Stancil1994}. 
Consequently, a comparison between \(\mathrm{H}_2^+\) and \(\mathrm{H}^-\) must include not only the different reactants in their formation reactions, but also the internal statistical structure of the molecular ion \cite{Short2021, Stancil1994}.

The atmospheric structures in which these equilibria must be evaluated are provided by model-atmosphere grids such as MARCS. 
The MARCS grid consists of one-dimensional LTE model atmospheres calculated under hydrostatic equilibrium, with mixing-length convection and either plane-parallel or spherical geometry depending on the stellar parameters \cite{Gustafsson2008,Plez2008, Bonifacio2011}. 
The grid covers late-type stellar atmospheres over a broad range of effective temperatures, surface gravities, metallicities, and abundance patterns, and provides thermodynamic quantities such as temperature, gas pressure, electron pressure, density, and partial pressures of chemical species as functions of atmospheric depth \cite{Gustafsson2008}. 
The continuing development of large atmosphere-model databases, including modern ATLAS9-based grids, further demonstrates the importance of well-sampled stellar parameter spaces for spectral synthesis, abundance work, and photometric transformations \cite{Bonifacio2011, Mucciarelli2025, Piskunov2004, Howarth2010}. 
For the present problem, the essential feature of MARCS is that it provides the layer-by-layer partial pressures of the relevant hydrogen species, allowing the abundance ratio of \(\mathrm{H}_2^+\) and \(\mathrm{H}^-\) to be evaluated directly within the local atmospheric structure.

Previous studies have included \(\mathrm{H}^-\), \(\mathrm{H}_2\), \(\mathrm{H}_2^+\), and related hydrogen species in broader chemical-equilibrium and opacity calculations. 
Work on \(\mathrm{H}^-\) has mainly emphasised its role as a dominant continuum opacity source and the validity of its LTE treatment in late-type atmospheres \cite{BarklemAmarsi2024}. 
Work on \(\mathrm{H}_2^+\) has focused primarily on photodissociation cross-sections, free--free absorption, and total absorption coefficients in hydrogen plasmas \cite{LebedevPresnyakov2002, Lebedev2003}. 
Molecular-equilibrium studies provide the partition-function and equilibrium-constant framework required for abundance calculations \cite{BarklemCollet2016}, while MARCS provides realistic depth-dependent model-atmosphere structures \cite{Gustafsson2008, Plez2008}. 
Nevertheless, these elements have not usually been combined to isolate the ratio
\begin{equation}
    R_{\mathrm{H}_2^+/\mathrm{H}^-}
    =
    \frac{N(\mathrm{H}_2^+)}
         {N(\mathrm{H}^-)}
    \label{eq:intro_ratio_definition}
\end{equation}
as a systematic layer-by-layer diagnostic across the MARCS atmosphere grid. 
To the best of our knowledge, the dependence of this ratio on effective temperature, metallicity, surface gravity, and atmospheric depth has not been mapped in detail for the full MARCS grid.

A useful theoretical starting point is provided by compact Saha-type concentration relations \cite{Stancil1994, BarklemCollet2016, LambertPagel1968}. 
For the negative hydrogen ion, the equilibrium abundance may be written as
\begin{equation}
    N(\mathrm{H}^-)
    =
    K_{\mathrm{H}^-}(T)\,
    N_{\mathrm H}\,N_e ,
    \label{eq:intro_hminus_saha}
\end{equation}
where \(N_{\mathrm H}\) is the number density of neutral hydrogen, \(N_e\) is the electron number density, and \(K_{\mathrm{H}^-}(T)\) is the temperature-dependent equilibrium concentration coefficient \cite{LambertPagel1968, BarklemAmarsi2024}. 
For the molecular hydrogen ion, the corresponding expression is
\begin{equation}
    N(\mathrm{H}_2^+)
    =
    K_{\mathrm{H}_2^+}(T)\,
    N_{\mathrm H}\,N_{\mathrm{H}^+},
    \label{eq:intro_h2plus_saha}
\end{equation}
where \(N_{\mathrm{H}^+}\) is the proton number density and \(K_{\mathrm{H}_2^+}(T)\) includes the internal partition function \(Z_{\mathrm{H}_2^+}(T)\)  \cite{Stancil1994,Babb2014}. 
Taking the ratio of Eqs.~\eqref{eq:intro_hminus_saha} and \eqref{eq:intro_h2plus_saha} gives
\begin{equation}
    \frac{N(\mathrm{H}_2^+)}
         {N(\mathrm{H}^-)}
    =
    \frac{K_{\mathrm{H}_2^+}(T)}
         {K_{\mathrm{H}^-}(T)}
    \frac{N_{\mathrm{H}^+}}{N_e}.
    \label{eq:intro_ratio_saha}
\end{equation}
This expression separates two physically distinct factors. 
The coefficient ratio \(K_{\mathrm{H}_2^+}(T)/K_{\mathrm{H}^-}(T)\) is a temperature-dependent reference quantity controlled by the relevant binding-energy terms and, for \(\mathrm{H}_2^+\), by the internal partition function \cite{Stancil1994, Babb2014}. 
The full atmospheric abundance ratio also contains the local ionisation factor \(N_{\mathrm{H}^+}/N_e\), which varies with temperature, pressure, composition, and depth. 
Therefore, the actual atmospheric ratio cannot be inferred from temperature alone.

The MARCS partial pressures provide a direct route to the layer-by-layer abundance ratio. 
For a species \(i\), the number density is related to its partial pressure by
\begin{equation}
    N_i
    =
    \frac{P_i}{k_{\rm B}T},
    \label{eq:intro_ideal_gas}
\end{equation}
where \(P_i\) is the partial pressure, \(k_{\rm B}\) is the Boltzmann constant, and \(T\) is the local layer temperature. 
For two species in the same atmospheric layer, the factor \(k_{\rm B}T\) cancels. 
Thus, the logarithmic abundance ratio can be obtained directly from the MARCS partial pressures:
\begin{equation}
    \log_{10}
    \left[
    \frac{N(\mathrm{H}_2^+)}
         {N(\mathrm{H}^-)}
    \right]
    =
    \log_{10}P(\mathrm{H}_2^+)
    -
    \log_{10}P(\mathrm{H}^-).
    \label{eq:intro_partial_pressure_ratio}
\end{equation}
This relation makes the abundance ratio a transparent partial-pressure diagnostic of the local chemical state of each atmospheric layer \cite{Stanger1963}.

The aim of this work is to identify stellar-atmosphere regimes in which \(\mathrm{H}_2^+\) becomes comparable to, or exceeds, \(\mathrm{H}^-\) in equilibrium abundance. 
We combine compact Saha-type concentration coefficients with MARCS partial pressures and compute \(N(\mathrm{H}_2^+)/N(\mathrm{H}^-)\) layer by layer. 
The analysis is performed as a function of the global stellar parameters \(T_{\mathrm{eff}}\), \(\log g\), and \([\mathrm{Fe}/\mathrm{H}]\), as well as atmospheric depth. 
The present study is restricted to equilibrium abundances and partial-pressure ratios. 
It does not claim that a larger abundance of \(\mathrm{H}_2^+\) automatically implies a larger opacity contribution, because opacity also depends on wavelength-dependent cross-sections, bound--free and free--free radiative processes, and radiative-transfer effects.

The novelty of the present work is the systematic combination of these previously separate elements. 
We use compact equilibrium expressions, an explicitly evaluated \(\mathrm{H}_2^+\) internal partition function, and MARCS layer-by-layer partial pressures to map the \(\mathrm{H}_2^+/\mathrm{H}^-\) equilibrium abundance ratio across the MARCS stellar-atmosphere grid. 
Rather than considering an idealised single-temperature plasma or a single representative stellar atmosphere, we identify regions of parameter space and atmospheric depth where \(\mathrm{H}_2^+\) becomes chemically significant relative to \(\mathrm{H}^-\). 
This approach preserves the distinction between a temperature-dependent coefficient ratio and the full atmospheric abundance ratio, which includes the local ionisation balance.

The paper is organized as follows. 
Section~\ref{sec:theory} introduces the theoretical background, including the compact Saha-type relations for \(\mathrm{H}^-\) and \(\mathrm{H}_2^+\), the role of \(Z_{\mathrm{H}_2^+}(T)\), and the distinction between the coefficient ratio and the atmospheric abundance ratio. 
Section~\ref{sec:compact} presents the compact coefficient calculations as a temperature-dependent reference. 
Section~\ref{sec:methods} describes the MARCS atmosphere grid, the extracted partial pressures, the conversion to abundance ratios, and the data-quality checks. 
Section~\ref{sec:full_grid_results} presents the full MARCS-grid results, including the dependence of \(N(\mathrm{H}_2^+)/N(\mathrm{H}^-)\) on effective temperature, metallicity, surface gravity, and atmospheric depth. 
Section~\ref{sec:discussion} discusses the physical interpretation and the distinction between abundance ratio and opacity. 
The main conclusions are summarized in Section~\ref{sec:conclusions}.
 
\section{Theoretical background}
\label{sec:theory}

The Introduction defined the two equilibrium formation channels for 
\(\mathrm{H}^-\) and \(\mathrm{H}_2^+\) in reactions~\eqref{eq:intro_hminus_reaction} 
and \eqref{eq:intro_h2plus_reaction}, and introduced the corresponding 
compact concentration relations in Eqs.~\eqref{eq:intro_hminus_saha} and 
\eqref{eq:intro_h2plus_saha}. 
Here we specify the coefficient forms used in the numerical calculations and 
clarify the distinction between the temperature-dependent coefficient ratio 
and the full atmospheric abundance ratio. 
Throughout this section, \(T\) denotes the local gas temperature in kelvin, 
number densities are in \({\rm cm^{-3}}\), and the equilibrium concentration 
coefficients \(K_{\mathrm{H}^-}(T)\) and \(K_{\mathrm{H}_2^+}(T)\) have units of 
\({\rm cm^3}\).

For the negative hydrogen ion, the equilibrium coefficient used in this work is
\begin{equation}
    K_{\mathrm{H}^-}(T)
    =
    1.03533 \times 10^{-16}
    T^{-3/2}
    \exp\left(\frac{8749.81}{T}\right).
    \label{eq:khminus_compact}
\end{equation}
This coefficient corresponds to the equilibrium attachment of a free electron 
to neutral hydrogen and contains the translational Saha factor and the binding 
energy of \(\mathrm{H}^-\). 
It describes the temperature-dependent part of the relation between 
\(N(\mathrm{H}^-)\), \(N_{\mathrm H}\), and \(N_e\), while the actual abundance 
also depends on the local neutral-hydrogen and electron densities. 
The direct proportionality of \(N(\mathrm{H}^-)\) to \(N_e\) is especially 
important in stellar atmospheres, because the electron pressure is determined 
by the local ionisation balance and chemical composition.

For the molecular hydrogen ion, the coefficient is written as
\begin{equation}
    K_{\mathrm{H}_2^+}(T)
    =
    7.44063 \times 10^{-21}
    Z_{\mathrm{H}_2^+}(T)
    T^{-3/2}
    \exp\left(\frac{30763.58}{T}\right),
    \label{eq:kh2plus_compact}
\end{equation}
where \(Z_{\mathrm{H}_2^+}(T)\) \cite{Glushko} is the internal partition function of 
\(\mathrm{H}_2^+\). 
Unlike \(\mathrm{H}^-\), the molecular ion has a rovibrational structure that 
must be included in the equilibrium coefficient. 
The partition function therefore accounts for the temperature-dependent 
population of the bound internal states of \(\mathrm{H}_2^+\). 
For rovibrational levels labelled by \(v\) and \(J\), this contribution may be 
written as
\begin{equation}
    Z_{\mathrm{H}_2^+}(T)
    =
    \sum_{v,J}
    (2J+1)
    \exp\left(
        -\frac{hc\,\varepsilon_{vJ}}{k_{\rm B}T}
    \right),
    \label{eq:zh2plus_partition}
\end{equation}
where \(\varepsilon_{vJ}\) is the excitation energy of the rovibrational level 
relative to the lowest level, expressed in wavenumber units. 
The factor \(2J+1\) is the rotational degeneracy. 
This form follows the standard statistical-mechanical treatment of molecular 
partition functions, in which the accessible internal states are weighted by 
their degeneracies and Boltzmann factors \cite{BarklemCollet2016}. 
For \(\mathrm{H}_2^+\), the same physical idea is central to calculations of 
Boltzmann-averaged radiative processes involving many rovibrational states 
\cite{LebedevPresnyakov2002,Lebedev2003}.

Using the abundance-ratio relation already introduced in 
Eq.~\eqref{eq:intro_ratio_saha}, the coefficient ratio is
\begin{equation}
    \mathcal{K}(T)
    \equiv
    \frac{K_{\mathrm{H}_2^+}(T)}
         {K_{\mathrm{H}^-}(T)} .
    \label{eq:coefficient_ratio_definition}
\end{equation}
Substitution of Eqs.~\eqref{eq:khminus_compact} and 
\eqref{eq:kh2plus_compact} gives
\begin{equation}
    \mathcal{K}(T)
    =
    7.18670 \times 10^{-5}
    Z_{\mathrm{H}_2^+}(T)
    \exp\left(\frac{22013.77}{T}\right).
    \label{eq:coefficient_ratio_compact}
\end{equation}
The factors \(T^{-3/2}\) cancel because both equilibrium coefficients contain 
the same translational temperature dependence. 
The remaining temperature dependence is controlled by the difference between 
the exponential binding-energy terms and by the internal partition function of 
\(\mathrm{H}_2^+\). 
Thus, \(\mathcal{K}(T)\) is a temperature-dependent reference quantity: it 
compares the intrinsic equilibrium coefficients of the two formation channels, 
but it does not yet represent the full atmospheric abundance ratio.

The full atmospheric ratio is obtained by multiplying the coefficient ratio by 
the local ionisation factor:
\begin{equation}
    \frac{N(\mathrm{H}_2^+)}
         {N(\mathrm{H}^-)}
    =
    \mathcal{K}(T)
    \frac{N_{\mathrm{H}^+}}{N_e}.
    \label{eq:atmospheric_ratio_theory}
\end{equation}
This equation is the central theoretical relation used to interpret the later 
MARCS results. 
The coefficient ratio \(\mathcal{K}(T)\) depends only on the local temperature 
and on \(Z_{\mathrm{H}_2^+}(T)\), whereas the factor 
\(N_{\mathrm{H}^+}/N_e\) is a local atmospheric quantity. 
It depends on the hydrogen ionisation state, the electron density, the gas 
pressure, chemical composition, and atmospheric depth. 
Consequently, the compact coefficient ratio and the MARCS abundance ratio are 
physically related but not identical. 
This distinction is essential because a grid-averaged trend with 
\(T_{\mathrm{eff}}\) does not correspond to inserting \(T_{\mathrm{eff}}\) 
directly into the local equilibrium coefficients; the abundance ratio in each 
layer is controlled by the local thermodynamic state.

For interpretation, we use the logarithmic diagnostic
\begin{equation}
    \mathcal{R}
    =
    \log_{10}
    \left[
        \frac{N(\mathrm{H}_2^+)}
             {N(\mathrm{H}^-)}
    \right].
    \label{eq:log_ratio_diagnostic}
\end{equation}
The three possible regimes are
\[
    \mathcal{R}<0
    \quad \Longleftrightarrow \quad
    N(\mathrm{H}_2^+) < N(\mathrm{H}^-),
\]
\[
    \mathcal{R}=0
    \quad \Longleftrightarrow \quad
    N(\mathrm{H}_2^+) = N(\mathrm{H}^-),
\]
and
\[
    \mathcal{R}>0
    \quad \Longleftrightarrow \quad
    N(\mathrm{H}_2^+) > N(\mathrm{H}^-).
\]
This notation is used throughout the MARCS analysis because the ratio can vary 
over many orders of magnitude across the grid.

Finally, the ratio considered in this work is an equilibrium abundance or 
partial-pressure diagnostic. 
It should not be interpreted as a direct opacity ratio. 
Even when \(N(\mathrm{H}_2^+)>N(\mathrm{H}^-)\), the corresponding opacity 
contribution depends on wavelength-dependent cross-sections and on the relevant 
bound--free and free--free radiative processes. 
The theoretical role of this section is therefore to define the compact 
coefficient ratio, identify the additional atmospheric factor 
\(N_{\mathrm{H}^+}/N_e\), and establish the notation used for the subsequent 
layer-by-layer MARCS analysis.\\

\section{Compact formula calculations}
\label{sec:compact}

\subsection{Temperature range}
\label{subsec:compact_temperature_range}

The compact Saha-type concentration coefficients defined in Section~\ref{sec:theory} were evaluated as functions of the local gas temperature \(T\). 
The purpose of this calculation is to establish a temperature-dependent reference for the intrinsic equilibrium behaviour of \(\mathrm{H}^-\) and \(\mathrm{H}_2^+\), before the abundance ratio is evaluated from the depth-dependent partial pressures of MARCS atmosphere models.

The coefficients considered in this section are \(K_{\mathrm{H}^-}(T)\), \(K_{\mathrm{H}_2^+}(T)\), and the coefficient ratio
\[
    \mathcal{K}(T)
    =
    \frac{K_{\mathrm{H}_2^+}(T)}
         {K_{\mathrm{H}^-}(T)} .
\]
The compact expressions for \(K_{\mathrm{H}^-}(T)\), \(K_{\mathrm{H}_2^+}(T)\), and \(\mathcal{K}(T)\) are given in Eqs.~\eqref{eq:khminus_compact}--\eqref{eq:coefficient_ratio_compact}. 
Here \(T\) is expressed in kelvin, and the equilibrium concentration coefficients are expressed in \({\rm cm^3}\).

The coefficients were evaluated over the interval
\[
    500 \leq T \leq 20\,000~\mathrm{K}.
\]
The main validated temperature interval is \(500\)--\(15\,000~\mathrm{K}\), while the extension to \(20\,000~\mathrm{K}\) is used only to illustrate the high-temperature behaviour of the compact analytical expressions. 
Results above \(15\,000~\mathrm{K}\) are therefore used as a qualitative continuation of the trend and should not be assigned the same validation status as the principal temperature range.

These compact calculations do not represent a complete stellar atmosphere. 
They include the temperature dependence of the analytical equilibrium coefficients and the internal partition function \(Z_{\mathrm{H}_2^+}(T)\), but they do not include the local electron density, proton density, gas pressure, electron pressure, metallicity, or optical-depth structure of a stellar atmosphere. 
They should therefore be interpreted as a theoretical baseline for the later MARCS layer-by-layer analysis.

\subsection{Results of compact coefficients}
\label{subsec:compact_results}

\begin{figure*}[htbp]
    \centering
    \includegraphics[width=\textwidth]{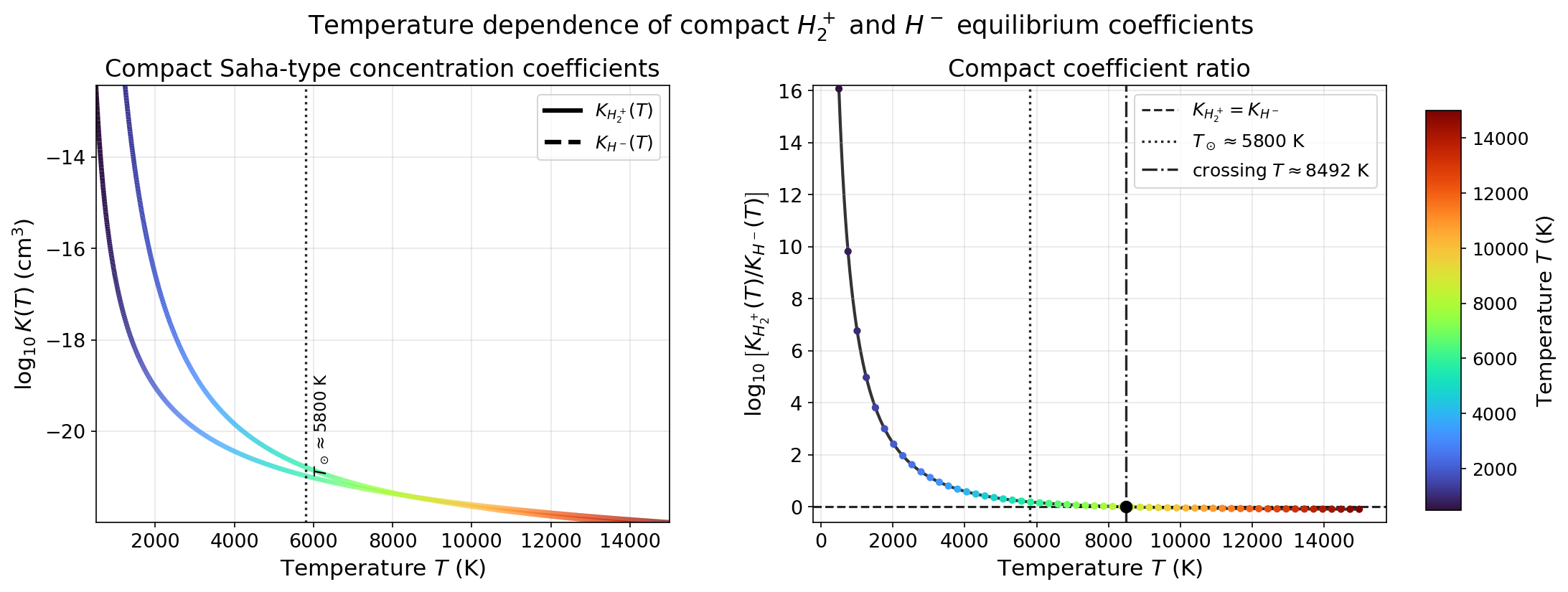}
    \caption{
    Temperature dependence of the compact Saha-type equilibrium concentration coefficients for \(\mathrm{H}_2^+\) and \(\mathrm{H}^-\). 
    The left panel shows \(\log_{10}K_{\mathrm{H}_2^+}(T)\) and \(\log_{10}K_{\mathrm{H}^-}(T)\), with coefficients expressed in \({\rm cm^3}\). 
    The solid curve corresponds to \(K_{\mathrm{H}_2^+}(T)\), and the dashed curve corresponds to \(K_{\mathrm{H}^-}(T)\). 
    The dotted vertical line marks the solar reference temperature \(T_\odot\simeq5800~\mathrm{K}\). 
    The right panel shows the compact coefficient ratio \(\log_{10}[K_{\mathrm{H}_2^+}(T)/K_{\mathrm{H}^-}(T)]\). 
    The horizontal dashed line marks equality of the two coefficients, and the dash-dotted vertical line marks the crossing temperature \(T\simeq8492~\mathrm{K}\). 
    This figure shows only the compact temperature-dependent coefficient ratio; the full atmospheric abundance ratio also depends on the local ionisation factor \(N_{\mathrm{H}^+}/N_e\).
    }
    \label{fig:compact_coefficients}
\end{figure*}

Figure~\ref{fig:compact_coefficients} shows the temperature dependence of the compact equilibrium coefficients and their ratio. 
The left panel presents \(\log_{10}K_{\mathrm{H}_2^+}(T)\) and \(\log_{10}K_{\mathrm{H}^-}(T)\) as functions of temperature. 
Both coefficients decrease with increasing \(T\) over the plotted range. 
This decrease reflects the reduced thermodynamic favourability of bound hydrogenic species relative to their separated reactants as the gas temperature increases. 
The two coefficients, however, do not have identical temperature dependences. 
For \(\mathrm{H}^-\), the behaviour is governed by the electron-affinity term and the common translational Saha factor. 
For \(\mathrm{H}_2^+\), the temperature dependence also includes the molecular binding-energy term and the internal partition function \(Z_{\mathrm{H}_2^+}(T)\), which accounts for the thermally populated internal states of the molecular ion.

The right panel of Fig.~\ref{fig:compact_coefficients} shows the logarithmic compact coefficient ratio $\log_{10}\!\left[K_{\mathrm{H}_2^+}(T)/K_{\mathrm{H}^-}(T)\right]$.
This ratio decreases strongly with increasing temperature. 
Under the simplifying reference condition $N_{\mathrm{H}^+}/N_e = 1$, the compact coefficient ratio becomes equal to the abundance ratio,
\[
    \frac{N(\mathrm{H}_2^+)}
         {N(\mathrm{H}^-)}
    =
    \frac{K_{\mathrm{H}_2^+}(T)}
         {K_{\mathrm{H}^-}(T)} .
\]
In this restricted reference case, equality between the two abundances corresponds to
\[
    \log_{10}
    \left[
    \frac{K_{\mathrm{H}_2^+}(T)}
         {K_{\mathrm{H}^-}(T)}
    \right]
    =0 .
\]
The compact calculation gives this equality at approximately $T \simeq 8492~\mathrm{K}$.

Below this reference temperature, the compact coefficient ratio is greater than unity; above it, the ratio is below unity. 
This value should not be interpreted as a universal transition temperature for stellar atmospheres. 
It is only the crossing temperature of the temperature-dependent coefficient ratio under the assumption \(N_{\mathrm{H}^+}/N_e=1\).

The compact coefficient ratio is therefore a useful temperature-dependent reference, but it is not the same as the actual stellar-atmosphere abundance ratio. 
As established in Section~\ref{sec:theory}, the full ratio is
\begin{equation}
    \frac{N(\mathrm{H}_2^+)}
         {N(\mathrm{H}^-)}
    =
    \frac{K_{\mathrm{H}_2^+}(T)}
         {K_{\mathrm{H}^-}(T)}
    \frac{N_{\mathrm{H}^+}}{N_e}.
    \label{eq:atmospheric_abundance_ratio}
\end{equation}
The additional factor \(N_{\mathrm{H}^+}/N_e\) is determined locally by the atmospheric ionisation balance and varies with depth-dependent thermodynamic conditions. 
Consequently, the compact coefficient ratio may decrease with \(T\), while the MARCS abundance ratio can show a different trend when organised by the global stellar parameter \(T_{\mathrm{eff}}\).

\section{MARCS data and extraction method}
\label{sec:methods}

\subsection{MARCS atmosphere grid}
\label{subsec:marcs_grid}

The stellar-atmosphere data used in this work were taken from the MARCS model-atmosphere grid. 
MARCS models provide one-dimensional stellar-atmosphere structures calculated under LTE, hydrostatic equilibrium, and either plane-parallel or spherical geometry, together with thermodynamic quantities and chemical-equilibrium partial pressures as functions of atmospheric depth \cite{Gustafsson2008, Plez2008}. 
These properties make the grid suitable for a layer-by-layer evaluation of the equilibrium abundance ratio \(N(\mathrm{H}_2^+)/N(\mathrm{H}^-)\).

The global MARCS database processed in this study contains \(51\,994\) standard model files. 
Each model has \(56\) atmospheric layers, giving \(2\,911\,664\) individual layer entries. 
The individual layer, rather than the model as a single global object, is the basic unit of the abundance-ratio calculation. 
Each layer has its own local temperature, pressure structure, electron pressure, and chemical-equilibrium partial pressures.

Both MARCS geometries were included. 
In the model filenames, \(p\) denotes plane-parallel geometry and \(s\) denotes spherical geometry. 
The two geometries are retained as part of the model metadata. 
No assumption is made that the plane-parallel and spherical subsets cover identical regions of parameter space. 
The standard MARCS solar model \texttt{sun.mod} was also processed as a reference case, but it was treated separately and excluded from the global \(T_{\mathrm{eff}}\)-dependent and \(T_{\mathrm{eff}}\)--\([\mathrm{Fe}/\mathrm{H}]\) grid statistics.

\begin{figure}[htbp]
    \centering
    \includegraphics[width=\textwidth]{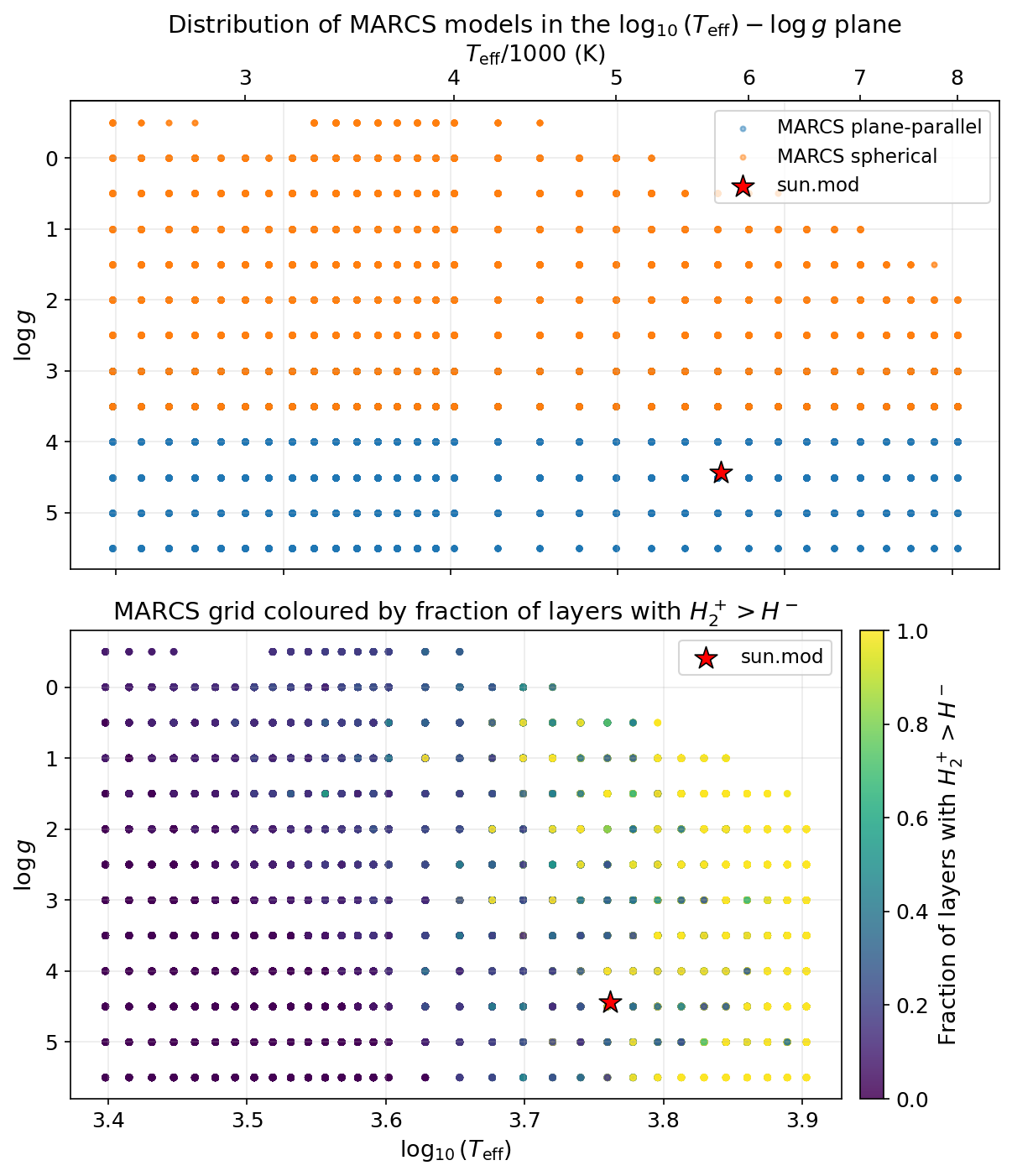}
    \caption{
    Distribution of the processed MARCS model atmospheres in the 
    \(\log_{10}(T_{\mathrm{eff}})\)--\(\log g\) plane. 
    The upper panel shows the grid coverage: blue points denote plane-parallel 
    MARCS models, orange points denote spherical models, and the red star marks 
    the solar reference model \texttt{sun.mod}. 
    Plane-parallel models mainly occupy the high-gravity part of the grid, 
    while spherical models cover lower-gravity regimes. 
    The lower panel shows the same parameter space coloured by the fraction 
    \(F=N_{\rm layers}(\mathrm{H}_2^+>\mathrm{H}^-)/N_{\rm layers,total}\). 
    Values near \(F=0\) indicate models in which \(\mathrm{H}^-\) is more abundant 
    than \(\mathrm{H}_2^+\) in nearly all layers, whereas values near \(F=1\) 
    indicate models in which \(\mathrm{H}_2^+\) exceeds \(\mathrm{H}^-\) in most 
    or all layers. 
    The colour-coded diagnostic represents an equilibrium abundance or 
    partial-pressure ratio and should not be interpreted as an opacity measure.
    }
    \label{fig:marcs_grid_distribution}
\end{figure}

Figure~\ref{fig:marcs_grid_distribution} shows the distribution of the processed MARCS models in the \(\log_{10}(T_{\mathrm{eff}})\)--\(\log g\) plane. 
The upper panel shows the parameter-space coverage and distinguishes plane-parallel and spherical models. 
The solar reference model \texttt{sun.mod} is marked separately near \(T_{\mathrm{eff}}\simeq5777~\mathrm{K}\) and \(\log g\simeq4.44\). 
The lower panel shows the same parameter plane coloured by the fraction of layers in which \(N(\mathrm{H}_2^+)>N(\mathrm{H}^-)\). 
This panel is included as a grid diagnostic: it illustrates where the processed models lie and how the layer-fraction diagnostic is distributed across the model grid. 
The detailed numerical trends are analysed in Section~\ref{sec:full_grid_results}.

\subsection{Extracted atmospheric quantities}
\label{subsec:extracted_quantities}

For each atmospheric layer, the extraction procedure retained the model metadata and the local atmospheric quantities needed to evaluate the \(\mathrm{H}_2^+/\mathrm{H}^-\) abundance ratio. 
The model-level quantities are \(T_{\mathrm{eff}}\), \(\log g\), \([\mathrm{Fe}/\mathrm{H}]\), and geometry. 

The layer-level quantities are the atmospheric layer index, local layer
temperature \(T_{\mathrm{layer}}\), Rosseland optical-depth coordinate 
\(\log_{10}\tau_{\mathrm{Ross}}\), electron pressure \(P_e\), gas pressure 
\(P_{\mathrm g}\), and the logarithmic partial pressures 
\(\log_{10}P(\mathrm{H})\), \(\log_{10}P(\mathrm{H}^-)\), 
\(\log_{10}P(\mathrm{H}_2)\), and \(\log_{10}P(\mathrm{H}_2^+)\).

The two quantities directly required for the abundance-ratio calculation are \(\log_{10}P(\mathrm{H}_2^+)\) and \(\log_{10}P(\mathrm{H}^-)\). 
The remaining quantities provide the thermodynamic and chemical context for each layer and are used for grouping, validation, and interpretation in later sections. 

The Rosseland optical-depth coordinate \(\log_{10}\tau_{\mathrm{Ross}}\) was adopted as the atmospheric depth coordinate for the layer-by-layer analysis.

\subsection{Calculation of the abundance ratio}
\label{subsec:abundance_ratio_calculation}

The conversion between partial pressure and number density was introduced in Eq.~\eqref{eq:intro_ideal_gas}. 
For two species evaluated in the same atmospheric layer, the factor \(k_{\rm B}T_{\mathrm{layer}}\) cancels in the ratio. 
Therefore, the logarithmic abundance ratio can be calculated directly from the difference between the two logarithmic partial pressures, as given in Eq.~\eqref{eq:intro_partial_pressure_ratio}. 
For each layer \(j\), the stored diagnostic is
\begin{equation}
    \mathcal{R}_j
    =
    \log_{10}
    \left[
    \frac{N_j(\mathrm{H}_2^+)}
         {N_j(\mathrm{H}^-)}
    \right]
    =
    \log_{10}P_j(\mathrm{H}_2^+)
    -
    \log_{10}P_j(\mathrm{H}^-).
    \label{eq:layer_log_ratio}
\end{equation}
This expression is evaluated independently for every atmospheric layer. 
The interpretation follows the notation established in Section~\ref{sec:theory}: \(\mathcal{R}_j<0\) means that \(\mathrm{H}^-\) is more abundant in that layer, \(\mathcal{R}_j=0\) corresponds to equality, and \(\mathcal{R}_j>0\) means that \(\mathrm{H}_2^+\) is more abundant.

This MARCS-based ratio differs from the compact coefficient ratio discussed in Section~\ref{sec:compact}. 
The compact ratio provides a temperature-dependent reference, while the MARCS diagnostic uses the equilibrium partial pressures already determined by the local atmospheric structure of each model layer.

\subsection{Statistical diagnostics and grouping}
\label{subsec:statistical_diagnostics}

Two statistical diagnostics are used in the subsequent analysis. 
The first is the fraction of layers in a selected set for which \(\mathrm{H}_2^+\) exceeds \(\mathrm{H}^-\):
\begin{equation}
    F
    =
    \frac{
    N_{\mathrm{layers}}
    \left[
    N(\mathrm{H}_2^+) > N(\mathrm{H}^-)
    \right]
    }
    {N_{\mathrm{layers,total}}}.
    \label{eq:fraction_layers_h2plus_gt_hminus}
\end{equation}
This quantity is dimensionless. 
A value \(F=0\) means that no layer in the selected set satisfies \(N(\mathrm{H}_2^+)>N(\mathrm{H}^-)\), while \(F=1\) means that all layers in the selected set satisfy this condition. 
Intermediate values measure the fraction of atmospheric layers in which \(\mathrm{H}_2^+\) is more abundant than \(\mathrm{H}^-\).

The second diagnostic is the arithmetic mean of the layer-by-layer logarithmic ratios,
\begin{equation}
    \left\langle \mathcal{R} \right\rangle
    =
    \frac{1}{N_{\mathrm{layers,total}}}
    \sum_{j=1}^{N_{\mathrm{layers,total}}}
    \mathcal{R}_j .
    \label{eq:mean_log_ratio}
\end{equation}
This is the mean of the logarithmic ratios, not the logarithm of the mean linear ratio. 
Negative values indicate that \(\mathrm{H}^-\) is more abundant on average in logarithmic space, zero corresponds to equality on the adopted logarithmic scale, and positive values indicate that \(\mathrm{H}_2^+\) is more abundant on average in logarithmic space.

The diagnostics were calculated for individual models, selected subsets, and binned regions of the MARCS parameter space. 
For grid-level trends, the ratios were grouped by \(T_{\mathrm{eff}}\), \([\mathrm{Fe}/\mathrm{H}]\), the \(T_{\mathrm{eff}}\)--\([\mathrm{Fe}/\mathrm{H}]\) plane, and depth coordinate. 
Statistics were calculated from atmospheric layers rather than from model files alone. 

\subsection{Quality control}
\label{subsec:quality_control}

The extraction procedure was validated before the abundance-ratio diagnostics were constructed. 
All \(51\,994\) selected MARCS model files were processed successfully. 
No failed files were recorded, and no unresolved metadata entries remained after parsing. 
Each processed model contained exactly \(56\) atmospheric layers, giving the expected total of \(2\,911\,664\) extracted layer entries. 
The local temperature column was checked separately from the depth coordinate, and no unphysical layer temperatures were found.

The abundance-ratio calculation was also checked directly from the extracted partial pressures. 
For every layer, the stored value of \(\mathcal{R}_j\) was verified against the difference
\[
    \log_{10}P_j(\mathrm{H}_2^+)
    -
    \log_{10}P_j(\mathrm{H}^-).
\]
This confirms the internal consistency of the extracted MARCS partial pressures and the computed logarithmic abundance ratios.

\section{Full MARCS-grid results}
\label{sec:full_grid_results}

The validated MARCS layer database described in Section~\ref{sec:methods} was used to evaluate the equilibrium abundance ratio of \(\mathrm{H}_2^+\) and \(\mathrm{H}^-\) over the full processed grid. 
The abundance diagnostic is the logarithmic ratio \(\mathcal{R}_j\) defined in Eq.~\eqref{eq:layer_log_ratio}, and the main statistical quantities are the layer fraction \(F\) and the mean logarithmic ratio \(\langle \mathcal{R} \rangle\) defined in Eqs.~\eqref{eq:fraction_layers_h2plus_gt_hminus} and \eqref{eq:mean_log_ratio}. 
All results in this section refer to equilibrium abundance or partial-pressure ratios derived from MARCS layer-by-layer partial pressures. 
They should not be interpreted as opacity ratios.

The full processed dataset contains \(51\,994\) MARCS model atmospheres and \(2\,911\,664\) atmospheric layers. 
Among these layers, \(827\,350\) satisfy
\[
    N(\mathrm{H}_2^+) > N(\mathrm{H}^-).
\]
The corresponding global layer fraction is
\[
    F
    =
    \frac{827\,350}{2\,911\,664}
    \simeq 0.284 ,
\]
which means that \(\mathrm{H}_2^+\) exceeds \(\mathrm{H}^-\) in approximately \(28.4\%\) of all analysed atmospheric layers. 
This is a fraction of layers, not a fraction of models. 
The mean logarithmic abundance ratio over the full layer database is
\[
    \left\langle \mathcal{R} \right\rangle
    =
    -4.229 .
\]
Thus, \(\mathrm{H}^-\) remains dominant in the global logarithmic average, while \(\mathrm{H}_2^+\) becomes more abundant than \(\mathrm{H}^-\) in a substantial subset of individual layers. 
These two statements are not contradictory: \(\langle \mathcal{R}\rangle\) measures the average logarithmic separation over the entire grid, whereas \(F\) counts how often the local abundance ratio exceeds unity. 
The largest linear ratio found in the processed data is approximately
\[
    \frac{N(\mathrm{H}_2^+)}
         {N(\mathrm{H}^-)}
    \simeq 10 ,
\]
corresponding to \(\mathcal{R}\simeq1\), while the minimum logarithmic ratio is approximately \(\mathcal{R}\simeq-23.94\). 
The wide range of \(\mathcal{R}\) shows that the relative abundance of the two species changes strongly across the MARCS parameter space and atmospheric depth.

\begin{figure}[htbp]
    \centering

    \begin{subfigure}{0.8\textwidth}
        \centering
        \includegraphics[width=\textwidth]{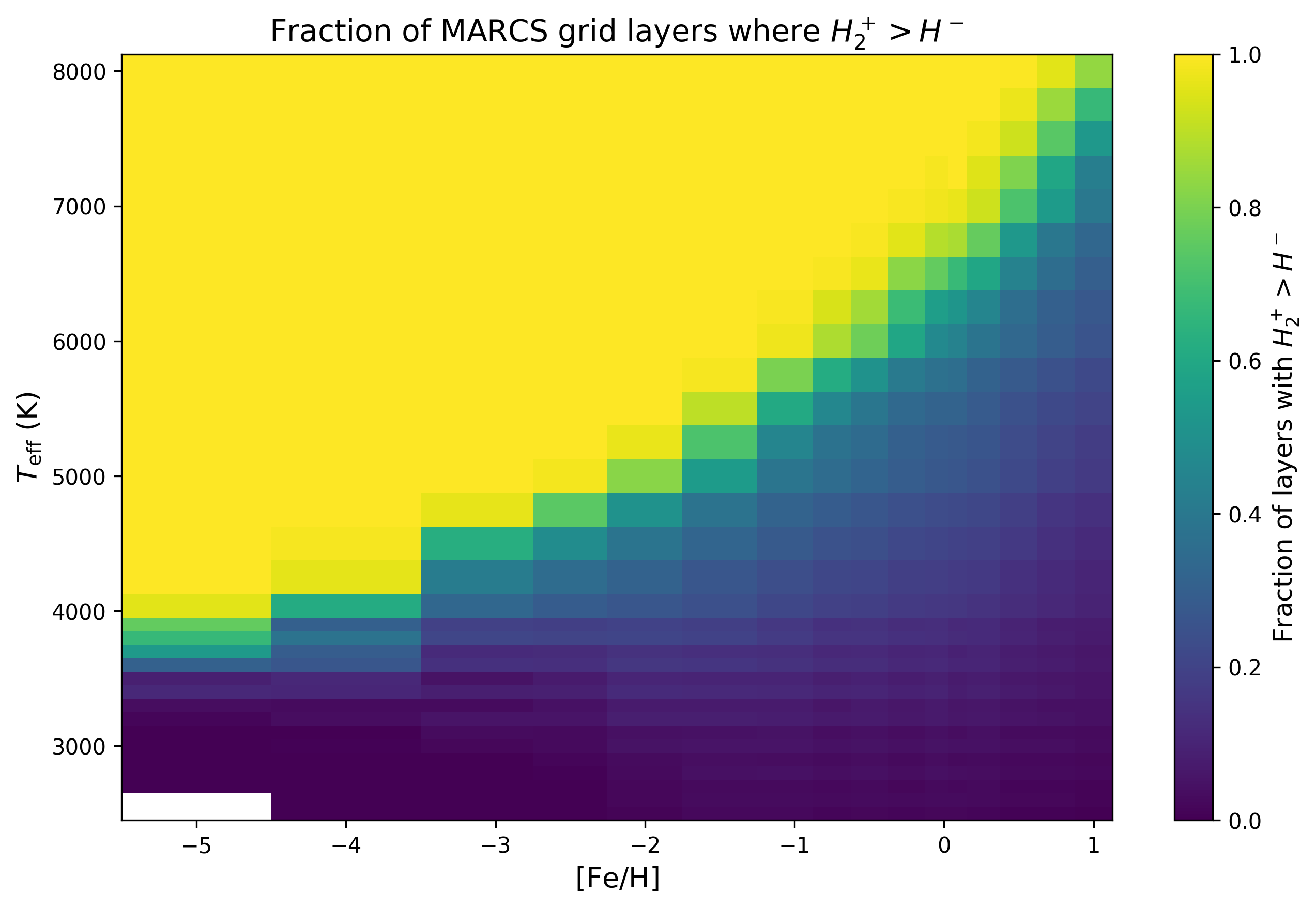}
        \caption{
        Fraction of MARCS atmospheric layers in which 
        \(N(\mathrm{H}_2^+) > N(\mathrm{H}^-)\) in the 
        \(T_{\mathrm{eff}}\)--\([\mathrm{Fe}/\mathrm{H}]\) plane.
        The colour scale gives 
        \(F=N_{\rm layers}[\mathrm{H}_2^+>\mathrm{H}^-]/N_{\rm layers,total}\).
        }
        \label{fig:heatmap_fraction}
    \end{subfigure}

    \vspace{0.5cm}

    \begin{subfigure}{0.8\textwidth}
        \centering
        \includegraphics[width=\textwidth]{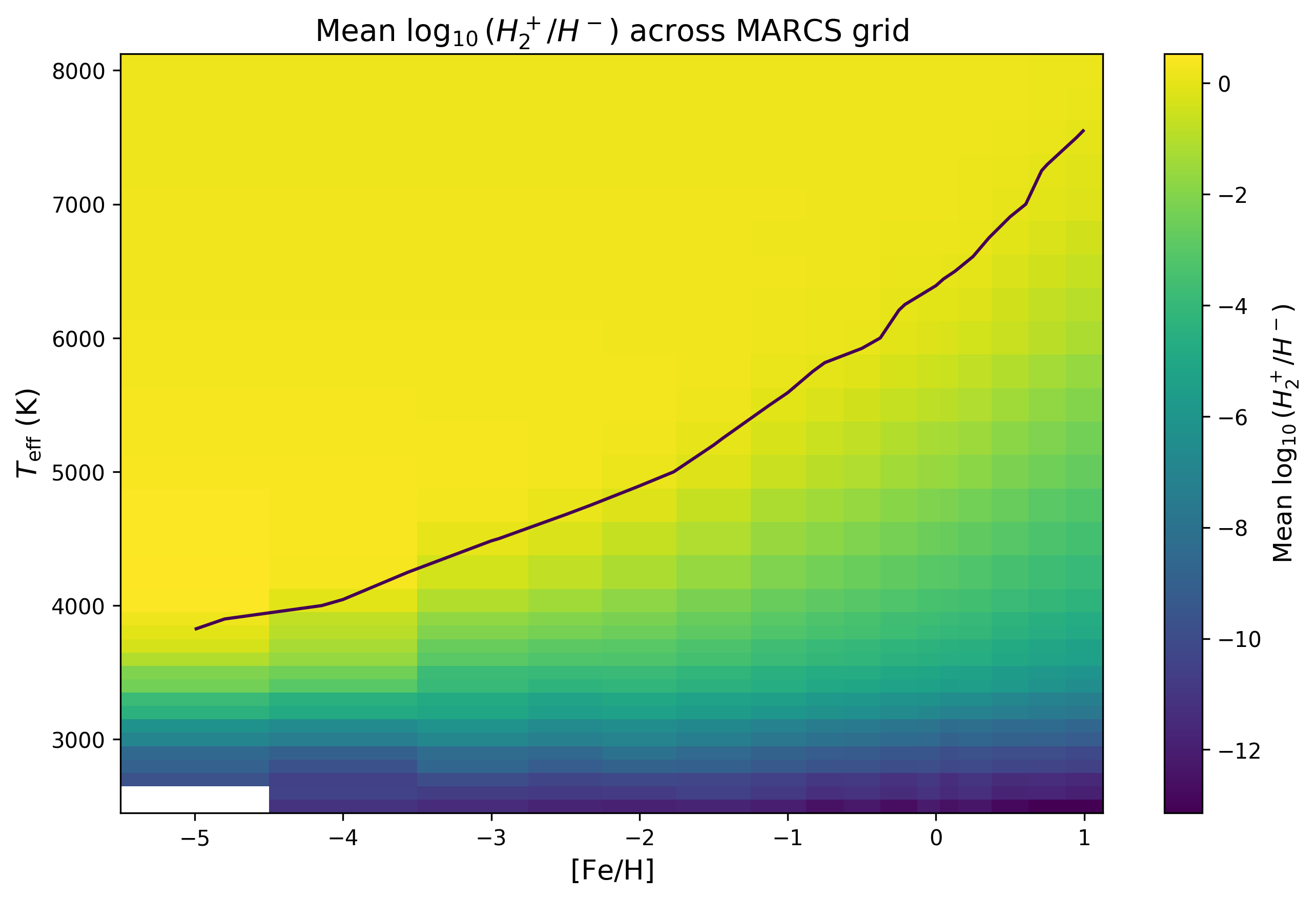}
        \caption{
        Mean logarithmic abundance ratio 
        \(\langle \log_{10}[N(\mathrm{H}_2^+)/N(\mathrm{H}^-)] \rangle\) 
        in the \(T_{\mathrm{eff}}\)--\([\mathrm{Fe}/\mathrm{H}]\) plane.
        The contour marks the approximate equality condition 
        \(\langle \mathcal{R} \rangle=0\).
        }
        \label{fig:heatmap_mean_log}
    \end{subfigure}

    \caption{
    MARCS-grid heatmaps of the equilibrium abundance ratio of 
    \(\mathrm{H}_2^+\) and \(\mathrm{H}^-\). 
    Panel (a) shows the fraction of atmospheric layers where 
    \(\mathrm{H}_2^+\) exceeds \(\mathrm{H}^-\). 
    Panel (b) shows the mean logarithmic abundance ratio in the same 
    \(T_{\mathrm{eff}}\)--\([\mathrm{Fe}/\mathrm{H}]\) plane. 
    Both diagnostics indicate that \(\mathrm{H}_2^+\) becomes more 
    chemically significant toward higher \(T_{\mathrm{eff}}\) and lower 
    metallicity.
    }
    \label{fig:teff_feh_heatmaps}
\end{figure}

The dependence of the abundance ratio on effective temperature and metallicity is summarized in Figs.~\ref{fig:heatmap_fraction} and \ref{fig:heatmap_mean_log}. 
Figure~\ref{fig:heatmap_fraction} shows the fraction \(F\) in the \(T_{\mathrm{eff}}\)--\([\mathrm{Fe}/\mathrm{H}]\) plane. 
The colour scale ranges from \(F=0\), where no layers in the bin satisfy \(N(\mathrm{H}_2^+)>N(\mathrm{H}^-)\), to \(F=1\), where this condition is satisfied in all layers of the bin. 
At low effective temperatures, especially below about \(3500\)--\(4000~\mathrm{K}\), \(F\) remains small over most metallicities, indicating that \(\mathrm{H}^-\) is more abundant in most atmospheric layers. 
As \(T_{\mathrm{eff}}\) increases, \(F\) rises, and large regions of the grid approach \(F\simeq1\). 
The transition from \(\mathrm{H}^-\)-dominated to \(\mathrm{H}_2^+\)-significant regimes is also metallicity dependent. 
At a given \(T_{\mathrm{eff}}\), metal-poor models generally show larger values of \(F\), while metal-rich models require higher \(T_{\mathrm{eff}}\) before \(\mathrm{H}_2^+\) exceeds \(\mathrm{H}^-\) in a large fraction of layers. 
The blank region in the lower-left part of the heatmap corresponds to parameter combinations not represented in the processed MARCS grid.

Figure~\ref{fig:heatmap_mean_log} presents the binned mean logarithmic abundance ratio,
\[
    \left\langle
    \log_{10}
    \left[
    \frac{N(\mathrm{H}_2^+)}
         {N(\mathrm{H}^-)}
    \right]
    \right\rangle ,
\]
over the same \(T_{\mathrm{eff}}\)--\([\mathrm{Fe}/\mathrm{H}]\) plane. 
Negative values indicate that \(\mathrm{H}^-\) is more abundant on average in logarithmic space, values close to zero indicate comparable abundances, and positive values indicate that \(\mathrm{H}_2^+\) is more abundant on the adopted binned mean diagnostic. 
The mean logarithmic ratio increases with increasing \(T_{\mathrm{eff}}\), moving from strongly negative values in cooler models toward values close to zero or above zero in hotter models. 
The metallicity dependence follows the same direction as in Fig.~\ref{fig:heatmap_fraction}: at fixed \(T_{\mathrm{eff}}\), lower metallicity generally corresponds to a larger mean \(\mathrm{H}_2^+/\mathrm{H}^-\) ratio. 
The contour in Fig.~\ref{fig:heatmap_mean_log} marks the approximate equality condition
\[
    \left\langle \mathcal{R} \right\rangle = 0 .
\]
This contour should not be interpreted as a layer-by-layer equality condition for every model in the bin. 
It marks equality of the binned mean logarithmic diagnostic. 
The equality boundary shifts toward lower \(T_{\mathrm{eff}}\) as metallicity decreases, indicating that \(\mathrm{H}_2^+\) becomes chemically significant at lower effective temperatures in metal-poor MARCS atmospheres.

The two heatmaps therefore give a consistent result: the relative abundance of \(\mathrm{H}_2^+\) increases toward higher \(T_{\mathrm{eff}}\) and lower \([\mathrm{Fe}/\mathrm{H}]\). 
This trend is a grid-averaged atmospheric result, not a single-temperature Saha calculation. 
The compact coefficient ratio discussed in Section~\ref{sec:compact} depends only on local temperature and \(Z_{\mathrm{H}_2^+}(T)\), whereas the MARCS abundance ratio also contains the layer-dependent ionisation factor \(N_{\mathrm{H}^+}/N_e\). 
Consequently, the behaviour in Figs.~\ref{fig:heatmap_fraction} and \ref{fig:heatmap_mean_log} reflects the combined effect of the local temperature structure, gas pressure, electron pressure, metallicity, and hydrogen ionisation balance in the MARCS models.

\begin{figure}[htbp]
    \centering
    \includegraphics[width=\textwidth]{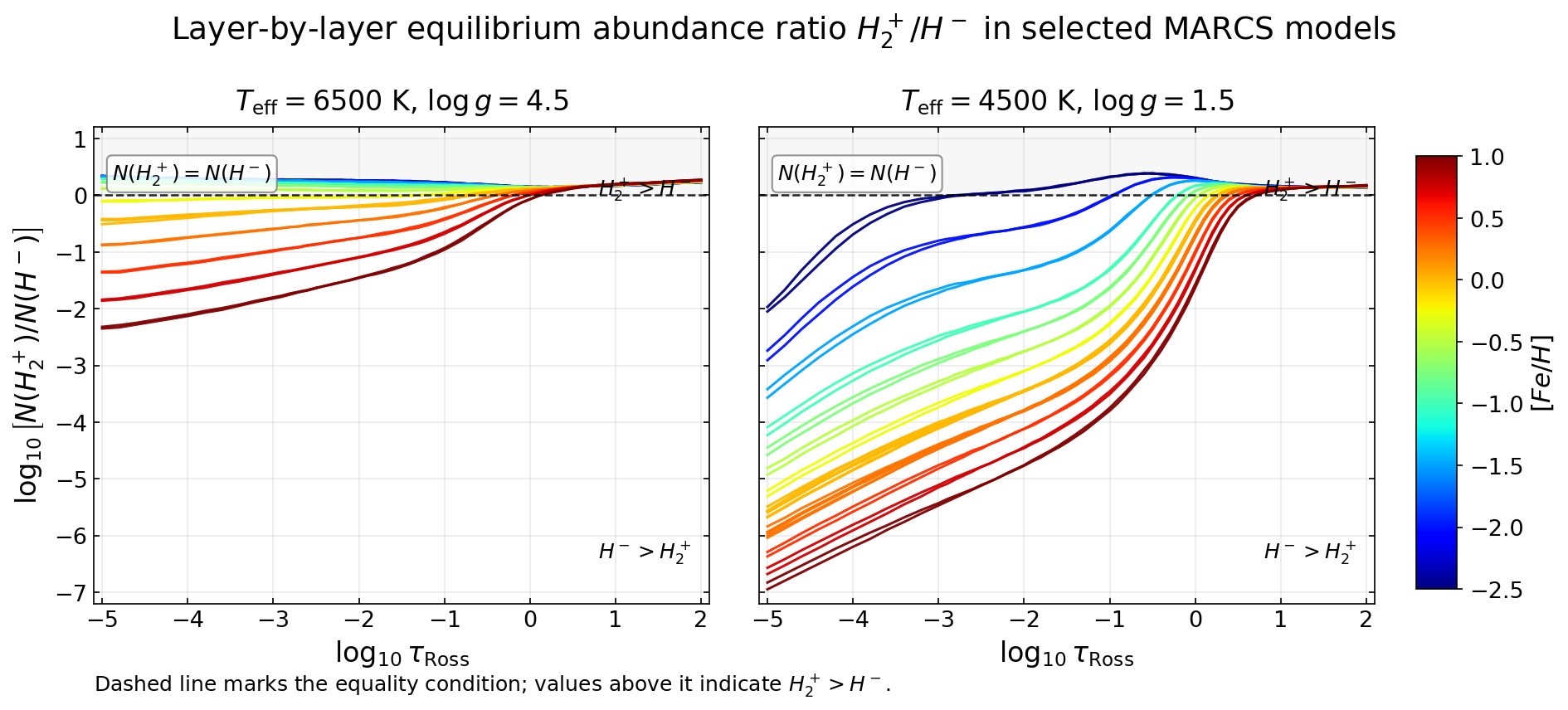}
    \caption{
    Layer-by-layer equilibrium abundance ratio 
    \(\log_{10}[N(\mathrm{H}_2^+)/N(\mathrm{H}^-)]\) as a function of 
    \(\log_{10}\tau_{\mathrm{Ross}}\) for selected MARCS models. 
    The left panel shows models with \(T_{\mathrm{eff}}=6500~\mathrm{K}\) and 
    \(\log g=4.5\), and the right panel shows models with 
    \(T_{\mathrm{eff}}=4500~\mathrm{K}\) and \(\log g=1.5\). 
    Each curve corresponds to one model with a different metallicity, shown by 
    the colour scale. 
    The dashed horizontal line marks the equality condition 
    \(N(\mathrm{H}_2^+)=N(\mathrm{H}^-)\). 
    Values above the line indicate layers where \(\mathrm{H}_2^+\) is more 
    abundant, while values below the line indicate \(\mathrm{H}^-\)-dominated 
    layers. 
    The figure is an equilibrium abundance diagnostic and should not be 
    interpreted as an opacity plot.
    }
    \label{fig:layer_profiles}
\end{figure}

The depth dependence of the abundance ratio is shown in Fig.~\ref{fig:layer_profiles} for two selected groups of MARCS models. 
The left panel corresponds to \(T_{\mathrm{eff}}=6500~\mathrm{K}\) and \(\log g=4.5\), while the right panel corresponds to \(T_{\mathrm{eff}}=4500~\mathrm{K}\) and \(\log g=1.5\). 
In each panel, the curves have the same \(T_{\mathrm{eff}}\) and \(\log g\), but different metallicities. 
The horizontal dashed line marks \(\mathcal{R}=0\), corresponding to \(N(\mathrm{H}_2^+)=N(\mathrm{H}^-)\). 
Values below the line indicate \(\mathrm{H}^-\)-dominated layers, while values above the line indicate layers in which \(\mathrm{H}_2^+\) exceeds \(\mathrm{H}^-\) in equilibrium abundance.

In the hotter high-gravity models, \(T_{\mathrm{eff}}=6500~\mathrm{K}\) and \(\log g=4.5\), the ratio is close to or above the equality line over a large part of the atmospheric structure, especially toward deeper layers. 
This indicates that \(\mathrm{H}_2^+\) becomes comparable to or exceeds \(\mathrm{H}^-\) in many layers of these models. 
In the cooler low-gravity models, \(T_{\mathrm{eff}}=4500~\mathrm{K}\) and \(\log g=1.5\), the upper atmosphere is more clearly \(\mathrm{H}^-\)-dominated, with negative \(\mathcal{R}\). 
However, the ratio increases with increasing optical depth and approaches or crosses the equality line near the photospheric and deeper layers. 
At fixed \(T_{\mathrm{eff}}\) and \(\log g\), metal-poor models generally show larger \(\mathrm{H}_2^+/\mathrm{H}^-\) ratios than metal-rich models in the upper and intermediate layers. 
The curves tend to converge in deeper layers, where local thermodynamic conditions become more similar in their effect on the ratio.

The layer-by-layer profiles demonstrate that the abundance ratio cannot be characterized by a single value for an entire model atmosphere. 
A given atmosphere may contain outer layers where \(\mathrm{H}^-\) is more abundant and deeper layers where \(\mathrm{H}_2^+\) becomes comparable to or exceeds \(\mathrm{H}^-\). 
The transition depends on \(T_{\mathrm{eff}}\), \(\log g\), metallicity, and the depth-dependent thermodynamic structure. 
The single solar reference model \texttt{sun.mod} was excluded from the global \(T_{\mathrm{eff}}\)--\([\mathrm{Fe}/\mathrm{H}]\) averages, but its behaviour is consistent with this general depth-dependent picture: \(\mathrm{H}^-\) dominates in upper layers, while \(\mathrm{H}_2^+\) approaches comparable abundance near the photospheric region and may exceed \(\mathrm{H}^-\) in deeper layers.

The full-grid results show that \(\mathrm{H}^-\) remains dominant in the global logarithmic average, but that \(\mathrm{H}_2^+\) exceeds \(\mathrm{H}^-\) in \(28.4\%\) of all analysed MARCS layers. 
The relative significance of \(\mathrm{H}_2^+\) increases toward higher \(T_{\mathrm{eff}}\), lower \([\mathrm{Fe}/\mathrm{H}]\), and deeper atmospheric layers. 
These trends identify regions where \(\mathrm{H}_2^+\) becomes chemically significant in equilibrium abundance. 
They do not imply that \(\mathrm{H}_2^+\) dominates the opacity, since opacity requires wavelength-dependent cross-sections and radiative-transfer calculations beyond the abundance-ratio diagnostic considered here.

In summary, the results presented above constitute a systematic equilibrium-abundance map of the relative importance of $H_2^+$ and $H_-$ in the MARCS parameter space.

\section{Discussion}
\label{sec:discussion}

The compact ratio, $K_{\mathrm{H}_2^+}(T)/K_{\mathrm{H}^-}(T)$, is a local temperature-dependent quantity determined by the adopted equilibrium coefficients and by the internal partition function \(Z_{\mathrm{H}_2^+}(T)\). 
It describes the intrinsic equilibrium tendency of the two formation channels at a given gas temperature. 
The abundance ratio evaluated in the MARCS models contains an additional atmospheric factor:
\[
    \frac{N(\mathrm{H}_2^+)}
         {N(\mathrm{H}^-)}
    =
    \frac{K_{\mathrm{H}_2^+}(T)}
         {K_{\mathrm{H}^-}(T)}
    \frac{N_{\mathrm{H}^+}}{N_e}.
    \label{eq:discussion_atmospheric_ratio}
\]
The factor \(N_{\mathrm{H}^+}/N_e\) depends on the local ionisation balance and on the electron reservoir of the atmospheric layer. 
Therefore, the compact coefficient ratio and the MARCS abundance ratio are physically related but not interchangeable.

This distinction explains why the temperature trend of the compact coefficient ratio and the grid-averaged MARCS trend with \(T_{\mathrm{eff}}\) need not be the same. 
The compact calculation uses the local gas temperature \(T\) and does not include the atmospheric stratification. 
The MARCS ratio is instead obtained from layer-by-layer partial pressures in models with depth-dependent \(T_{\mathrm{layer}}\), gas pressure, electron pressure, and chemical equilibrium. 
Moreover, \(T_{\mathrm{eff}}\) is a global property of a stellar model and should not be treated as the temperature inserted into each local equilibrium expression. 
A change in \(T_{\mathrm{eff}}\) changes the atmospheric structure as a whole and may change \(N_{\mathrm{H}^+}/N_e\) sufficiently to offset the temperature dependence of the compact coefficient ratio. 
The increasing relative importance of \(\mathrm{H}_2^+\) toward higher \(T_{\mathrm{eff}}\) in the MARCS grid should therefore be understood as a layer-averaged atmospheric result, not as a single-temperature Saha prediction.

The metallicity dependence follows from the same separation between coefficient and atmospheric ratios. 
The equilibrium abundance of the negative hydrogen ion is proportional to the electron density,
\[
    N(\mathrm{H}^-)
    =
    K_{\mathrm{H}^-}(T)\,
    N_{\mathrm H}\,N_e .
\]
Thus, changes in the electron reservoir directly affect \(N(\mathrm{H}^-)\). 
In cool and late-type stellar atmospheres, metals contribute to the electron population, and variations in metallicity can modify the electron pressure and the chemical-equilibrium partial pressures \cite{Gustafsson2008,BarklemAmarsi2024}. 
The shift of the equality boundary toward lower \(T_{\mathrm{eff}}\) at lower \([\mathrm{Fe}/\mathrm{H}]\) is consistent with reduced metal-derived electron availability contributing to a larger relative \(\mathrm{H}_2^+/\mathrm{H}^-\) abundance ratio. 
This interpretation should not be read as proof of a single isolated metallicity effect, because the heatmaps are averaged over model bins that may include different \(\log g\), geometries, and abundance patterns. 
Metallicity also affects the atmospheric pressure--temperature structure and the chemical equilibrium of the model atmosphere.

The layer-by-layer profiles show that atmospheric depth is another essential variable. 
Optical depth is not itself a chemical variable; moving through \(\log_{10}\tau_{\mathrm{Ross}}\) or \(\log_{10}\tau_5\) corresponds to sampling layers with different local thermodynamic conditions. 
The selected MARCS profiles show that the abundance ratio generally increases toward deeper layers. 
In the upper layers of solar-like and cooler models, \(\mathrm{H}^-\) is usually more abundant, whereas near the photospheric region and in deeper layers \(\mathrm{H}_2^+\) can become comparable to or exceed \(\mathrm{H}^-\) in equilibrium abundance. 
This behaviour follows from the combined depth dependence of \(K_{\mathrm{H}_2^+}(T_{\mathrm{layer}})/K_{\mathrm{H}^-}(T_{\mathrm{layer}})\) and \(N_{\mathrm{H}^+}/N_e\). 
It also shows why a single abundance ratio cannot represent an entire stellar atmosphere.

The atmospheric-structure profiles in Supplementary Fig.~\ref{fig:supp_atmospheric_structure} provide supporting context for this depth dependence. 
For the selected MARCS models, \(T_{\mathrm{layer}}\), \(P_{\mathrm g}\), and \(N_e\) vary systematically with \(\log_{10}\tau_{\mathrm{Ross}}\), and the profiles differ with metallicity at fixed \(T_{\mathrm{eff}}\) and \(\log g\). 
This supports the interpretation that the metallicity and depth trends in \(N(\mathrm{H}_2^+)/N(\mathrm{H}^-)\) arise from changes in the local atmospheric structure, rather than from the compact coefficient ratio alone. 
Supplementary Fig.~\ref{fig:supp_atmospheric_structure} is therefore used only as a thermodynamic diagnostic; it is not an abundance-ratio or opacity plot.

The present work also clarifies the role of \(\mathrm{H}_2^+\) relative to previous studies. 
The negative hydrogen ion is well established as a major source of continuous opacity in solar and late-type stellar atmospheres \cite{BarklemAmarsi2024}. 
The molecular hydrogen ion \(\mathrm{H}_2^+\) has been studied in the context of molecular structure and bound--free and free--free radiative processes in hydrogen plasmas \cite{LebedevPresnyakov2002,Lebedev2003}. 
The present analysis does not challenge the established opacity role of \(\mathrm{H}^-\). 
Its contribution is to quantify the equilibrium abundance ratio \(N(\mathrm{H}_2^+)/N(\mathrm{H}^-)\) layer by layer across the MARCS grid and to identify the atmospheric regimes where \(\mathrm{H}_2^+\) becomes chemically significant relative to \(\mathrm{H}^-\).

The abundance ratio must be distinguished from opacity. 
This study calculates equilibrium abundance and partial-pressure ratios, not wavelength-dependent opacity coefficients. 
Even when
\[
    N(\mathrm{H}_2^+) > N(\mathrm{H}^-),
\]
it does not follow that
\[
    \alpha_\nu(\mathrm{H}_2^+) >
    \alpha_\nu(\mathrm{H}^-).
\]
An opacity coefficient depends not only on number density, but also on cross sections, wavelength, radiative process, and radiative-transfer conditions. 
Schematically,
\[
    \alpha_{\nu,i}
    \propto
    N_i\,\sigma_{\nu,i},
\]
where \(N_i\) is the number density of species \(i\), and \(\sigma_{\nu,i}\) is the frequency-dependent cross section for the relevant process. 
The present maps therefore identify where \(\mathrm{H}_2^+\) is chemically important in equilibrium abundance, but they do not demonstrate opacity dominance. 
A direct opacity assessment would require wavelength-dependent \(\mathrm{H}_2^+\) and \(\mathrm{H}^-\) cross sections and radiative-transfer calculations.

The analysis is based on one-dimensional MARCS model atmospheres computed under LTE chemical equilibrium and hydrostatic assumptions \cite{Gustafsson2008,Plez2008}. 
The ratios are derived from equilibrium partial pressures and do not include time-dependent chemistry, non-LTE populations, or wavelength-dependent opacity calculations. 
The heatmaps are also grid-averaged diagnostics and may hide model-to-model and layer-to-layer dispersion within a given \(T_{\mathrm{eff}}\)--\([\mathrm{Fe}/\mathrm{H}]\) bin. 
Within these assumptions, the results show that the atmospheric abundance ratio is controlled by both the compact coefficient ratio and the local ionisation factor, and that \(\mathrm{H}_2^+\) becomes comparatively more significant in higher-\(T_{\mathrm{eff}}\), lower-metallicity, and deeper atmospheric layers.

\section{Conclusions}
\label{sec:conclusions}

We have investigated the equilibrium abundance ratio of the molecular hydrogen ion \(\mathrm{H}_2^+\) and the negative hydrogen ion \(\mathrm{H}^-\) in MARCS stellar-atmosphere models. The analysis combines compact Saha-type equilibrium concentration formulas with layer-by-layer partial pressures extracted from MARCS model atmospheres. For the two species, the adopted equilibrium relations are
\[
    N(\mathrm{H}^-)
    =
    K_{\mathrm{H}^-}(T)\,
    N_{\mathrm H}\,N_e,
\]
and
\[
    N(\mathrm{H}_2^+)
    =
    K_{\mathrm{H}_2^+}(T)\,
    N_{\mathrm H}\,N_{\mathrm{H}^+}.
\]
The internal partition function \(Z_{\mathrm{H}_2^+}(T)\) is included in the \(\mathrm{H}_2^+\) equilibrium coefficient, so that the molecular internal states enter the equilibrium abundance calculation.

A central conclusion of the theoretical analysis is that the compact coefficient ratio and the full atmospheric abundance ratio are related, but they are not identical quantities. The abundance ratio can be written as
\[
    \frac{N(\mathrm{H}_2^+)}
         {N(\mathrm{H}^-)}
    =
    \frac{K_{\mathrm{H}_2^+}(T)}
         {K_{\mathrm{H}^-}(T)}
    \frac{N_{\mathrm{H}^+}}{N_e}.
\]
The coefficient ratio \(K_{\mathrm{H}_2^+}(T)/K_{\mathrm{H}^-}(T)\) provides a temperature-dependent reference behaviour, while the full atmospheric ratio also contains the local ionisation factor \(N_{\mathrm{H}^+}/N_e\). This factor depends on the local chemical-equilibrium structure of each atmospheric layer and explains why the MARCS-grid behaviour cannot be inferred from the compact coefficient ratio alone.

The MARCS analysis includes \(51{,}994\) stellar-atmosphere models and 
\(2{,}911{,}664\) individual atmospheric layers. 
The equilibrium abundance ratio \(N(\mathrm{H}_2^+)/N(\mathrm{H}^-)\) was 
calculated separately for each layer from the tabulated MARCS partial pressures. 
The condition \(N(\mathrm{H}_2^+)>N(\mathrm{H}^-)\) is satisfied in 
\(827{,}350\) layers, corresponding to a layer fraction
\[
    F=\frac{827{,}350}{2{,}911{,}664}\simeq0.284 .
\]

Thus, \(\mathrm{H}_2^+\) exceeds \(\mathrm{H}^-\) in approximately \(28.4\%\) 
of all analysed layers. 
This value refers to the fraction of layers, not the fraction of MARCS models. 
Nevertheless, the global mean logarithmic ratio remains negative, indicating 
that \(\mathrm{H}^-\) is more abundant on average across the full database.
To our knowledge, the present work provides the first systematic mapping of the equilibrium abundance ratio 
$N(H_2^+)/N(H^-)$ in the MARCS stellar-atmosphere grid, thus identifying the atmospheric regimes in which $H_2^+$ 
becomes chemically significant relative to $H_-$. 

The grid-averaged results show that the relative abundance of 
\(\mathrm{H}_2^+\) increases toward higher \(T_{\mathrm{eff}}\) and lower 
\([\mathrm{Fe}/\mathrm{H}]\). Accordingly, the approximate equality boundary 
between \(\mathrm{H}_2^+\) and \(\mathrm{H}^-\) occurs at lower 
\(T_{\mathrm{eff}}\) in more metal-poor models. These trends represent averages 
over parameter bins that may include different values of \(\log g\), geometry, 
abundance pattern, and atmospheric structure. The layer-by-layer profiles also 
show that \(\mathrm{H}^-\) generally dominates in the outer atmosphere, whereas 
the ratio \(N(\mathrm{H}_2^+)/N(\mathrm{H}^-)\) increases toward the 
photospheric and deeper layers. In some deeper layers, \(\mathrm{H}_2^+\) 
becomes comparable to or more abundant than \(\mathrm{H}^-\).

The present results describe equilibrium abundance and partial-pressure ratios, not opacity. A larger \(N(\mathrm{H}_2^+)/N(\mathrm{H}^-)\) does not imply a larger \(\mathrm{H}_2^+\) opacity contribution, which also depends on wavelength-dependent cross-sections and radiative transfer. The conclusions are therefore limited to one-dimensional MARCS atmospheres under LTE chemical equilibrium. Within this framework, \(\mathrm{H}_2^+\) may require consideration in higher-\(T_{\mathrm{eff}}\), metal-poor, and deeper atmospheric layers.

\clearpage
\begin{appendices}

\section{Supplementary material}\label{secA1}
\label{sec:supplementary}

\begin{figure}[H]
    \centering
    \includegraphics[
        width=\textwidth,
        height=0.72\textheight,
        keepaspectratio
    ]{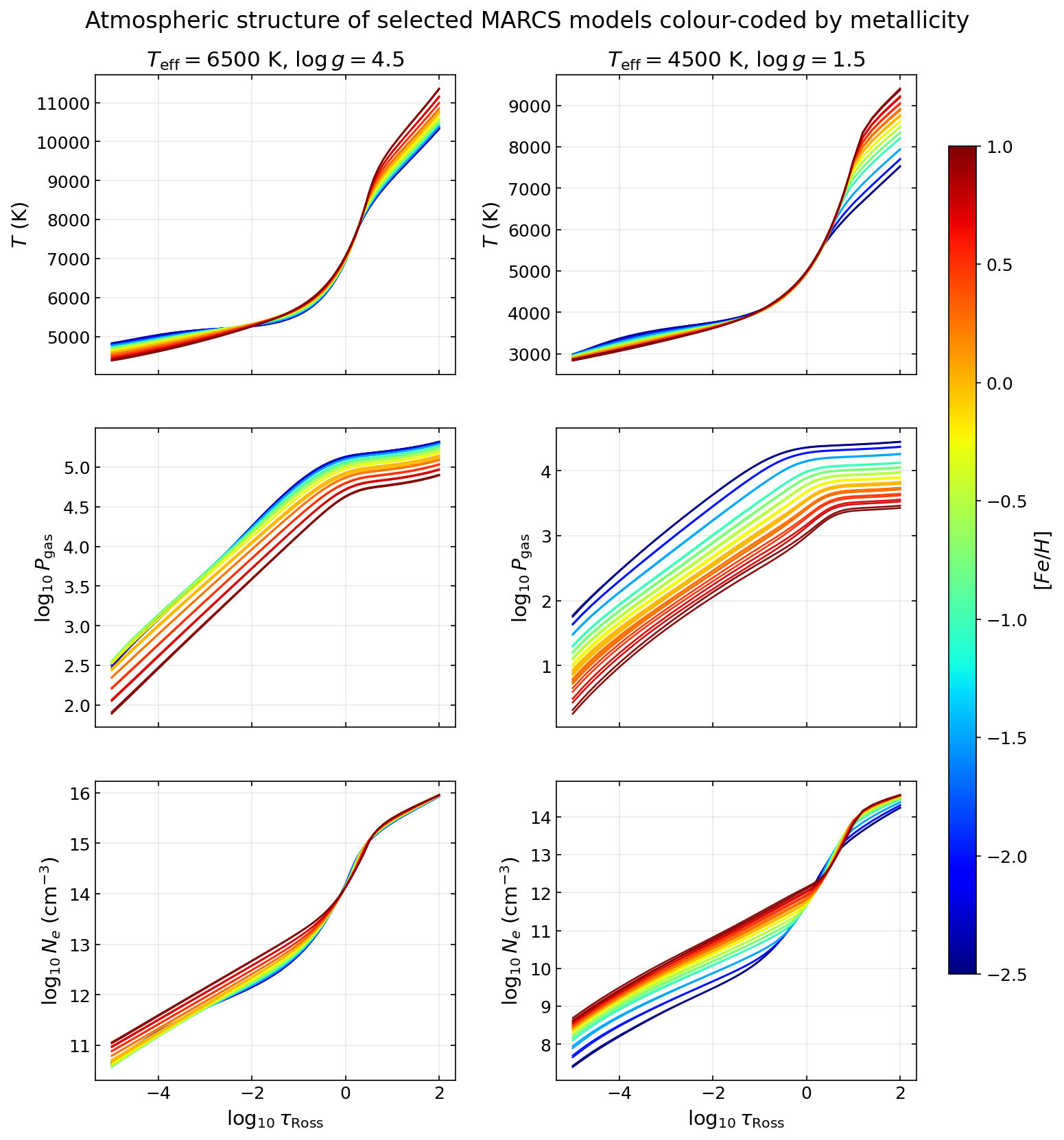}
    \caption{
    Atmospheric structure of selected MARCS models colour-coded by metallicity.
    The left column shows models with \(T_{\mathrm{eff}}=6500~\mathrm{K}\) and
    \(\log g=4.5\), while the right column shows models with
    \(T_{\mathrm{eff}}=4500~\mathrm{K}\) and \(\log g=1.5\).
    In each column, the models have the same \(T_{\mathrm{eff}}\) and
    \(\log g\), but different metallicities
    \([\mathrm{Fe}/\mathrm{H}]\), indicated by the colour scale.
    The upper row shows the local layer temperature
    \(T_{\mathrm{layer}}\), the middle row shows the gas pressure
    \(\log_{10}P_{\mathrm{g}}\), and the lower row shows the electron
    number density \(\log_{10}N_e\) in \(\mathrm{cm}^{-3}\), all as
    functions of \(\log_{10}\tau_{\mathrm{Ross}}\).
    }
    \label{fig:supp_atmospheric_structure}
\end{figure}

\end{appendices}

\section*{Declarations}

\hspace*{1.5em}\textbf{Funding}
No funding was received for conducting this study.

\textbf{Competing Interests}
The authors have no relevant financial or non-financial interests to disclose.

\textbf{Ethics approval and consent to participate}
Not applicable.

\textbf{Consent for publication}
Not applicable.

\textbf{Data availability}
The MARCS model atmosphere data analysed in this study are publicly available
from the MARCS model atmosphere database [https://marcs.astro.uu.se/data.html].
The derived data supporting the findings of this study are available from the
corresponding author upon reasonable request.

\textbf{Materials availability}
Not applicable.

\textbf{Code availability}
The computational codes used for the analysis are available from the
corresponding author upon reasonable request.

\textbf{Author contributions}
A.S.: Conceptualization, methodology, data analysis, numerical calculations, and writing--original draft. H.Q.: Interpretation of the results and writing--review and editing. All authors read and approved the final manuscript.


\end{document}